\documentclass[aps,pra,11pt,singlecolumn,tightenlines,nofootinbib]{revtex4}

\usepackage{color}
\usepackage{amsmath,amsfonts,amsthm}
\usepackage{graphicx}
\usepackage{bbm} 

\usepackage{tikz}
\usetikzlibrary{arrows,decorations.pathmorphing,backgrounds,positioning,fit}

\pgfrealjobname{KalaiC}

\usepackage[pdfpagelabels,pdftex,bookmarks,breaklinks]{hyperref}
\definecolor{deeppurple}{RGB}{100,0,120} 
\definecolor{darkgreen}{RGB}{0,150,0}
\definecolor{darkblue}{RGB}{0,0,130}
\hypersetup{colorlinks=true, linkcolor=deeppurple, citecolor=darkgreen, filecolor=red, urlcolor=darkblue}
\hypersetup{pdftitle={A Counterexample to Kalai's Conjecture C}}
\hypersetup{pdfauthor={Steven T. Flammia, Aram W. Harrow}}

\theoremstyle{remark}

\newcommand*{\C}{\mathbb{C}}

\newcommand*{\Eref}[1]{Eq.~(\ref{#1})}

\newcommand*{\Fref}[1]{Fig.~\ref{#1}}

\newcommand*{\tr}{\mathrm{tr}}

\newcommand*{\ket}[1]{|{#1}\rangle}

\newcommand*{\ketbra}[2]{|{#1}\rangle\!\langle{#2}|}

\newcommand*{\proj}[1]{\ketbra{#1}{#1}}

\DeclareMathOperator{\poly}{poly}

\newcommand{\cE}{\mathcal{E}}
\newcommand{\Sep}{\mathrm{Sep}}

\newcommand{\cN}{\mathcal{N}}

\def\ba#1\ea{\begin{align}#1\end{align}}
\def\ban#1\ean{\begin{align*}#1\end{align*}}

\begin{document}

\title{Counterexamples to Kalai's Conjecture C}

\author{Steven T.\ Flammia and Aram W.\ Harrow}
\affiliation{Department of Computer Science and Engineering, University of Washington, Seattle, WA}

\date{\today}

\begin{abstract}
We provide two simple counterexamples to Kalai's Conjecture C and discuss our perspective on the implications for the prospect of large-scale fault-tolerant quantum computation.
\end{abstract}

\maketitle

Let $\rho$ be a quantum state on the space $(\C^d)^{\otimes n}$ of $n$ qudits ($d$-dimensional quantum systems). Denote by $\rho_S$ the reduced state of $\rho$ onto some subset $S$ of the qudits. We partition $S$ further into two nonempty subsets of qudits $A$ and $A^c:=S\backslash A$ (thus $S$ must contain at least two qudits). Denote by $\Sep(A)$ the convex set of quantum states which are bipartite separable on $S$ across the bipartition defined by $A$ and $A^c$. Define the following function
\begin{align}
	\Delta(\rho_S) = \min_{A\subset S} \inf_{\sigma \in \Sep(A)} \|\rho_S - \sigma\|_\tr \,.
	\label{eq:Delta-def}
\end{align}
That is, $\Delta$ measures the distance in the trace norm between the
nearest biseparable state and $\rho_S$, across all bipartitions of
$S$.  Here $\|X\|_\tr := \frac{1}{2}\|X\|_1 := \frac{1}{2} \tr
\sqrt{X^\dag X}$.  Now define
\begin{align}
	K(\rho) = \sum_{S \subset [n]} \Delta(\rho_S) \,,
	\label{eq:K-def}
\end{align}
where $[n]:=\{1,\ldots,n\}$.
Kalai has made the following conjecture~\cite{Kalai-blog}, which he calls ``Conjecture C''.\\

\noindent
\textbf{Kalai's Conjecture C:}
\emph{For all states $\rho$ which have been efficiently prepared by a noisy quantum computer, there is a polynomial $P(n)$ such that $K(\rho) \le P(n)$.}\\

We remark that Kalai originally formulated his conjecture in terms of
qubits (2-level systems), but here we extend this to some
$d=O(1)$. The counterexample below requires $d=16$, but a simple
extension can improve it to $d=8$ (using a hexagonal grid instead of a
square grid).  What about $d=2$?  It is definitely possible to find
states on $n$ qubits for which $K(\rho)$ is large (e.g. random
states), but the tricky part is finding states with this property that
can clearly be created efficiently.  In this note, we focus on
considering states where it is (we hope) absolutely clear that there
is no in-principle barrier to constructing them.  However, we also
give an example where the elementary systems are qubits and the
physical plausibility is relatively uncontroversial.

We further remark that the formulation of what it means to ``be prepared by a noisy quantum computer'' was intentionally left somewhat vague by Kalai in order to stimulate further discussion about what constitutes a reasonable noise model for a quantum computer. Here we show that this conjecture is false for any noise model in which a constant trace distance from the set of separable states can be generated between a pair of nearest-neighbor qubits on a lattice. 

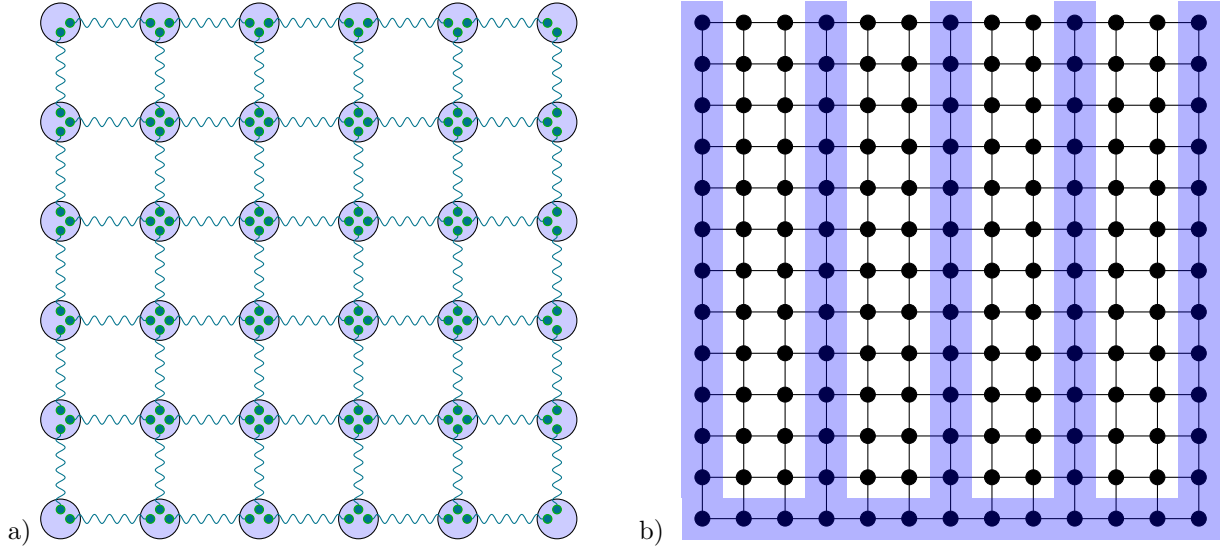
\begin{figure}[t!]
\centering
a)\beginpgfgraphicnamed{PEPS}
\begin{tikzpicture}
	[a/.style={circle, fill=green!45!blue, draw=green!70!black, scale=.35},
	b/.style={fill=none, decorate, 
	decoration={snake,amplitude=.6mm,segment length=2mm, 
	pre length=1.8, post length=.8}, green!45!blue},scale=2.2,
	c/.style={fill=blue,fill opacity = .2},scale=.6]

	\foreach \x in {0,1,...,5} \foreach \y in {0,1,...,5}{
		\filldraw[c, shift={(\x,\y)}] (0,0) circle (2mm);
	}

	\def\tick{.095}
	\foreach \x in {0,1,...,4}{
		\foreach \y in {0,1,...,4}{
			\filldraw[b] 
			(\x+\tick,\y) node [a] {} -- (\x+1-\tick,\y) node [a] {};
			\filldraw[b] 
			(\x,\y+\tick) node [a] {} -- (\x,\y+1-\tick) node [a] {};
		}
		\filldraw[b] 
		(\x+\tick,5) node [a] {} -- (\x+1-\tick,5) node [a] {};
		\filldraw[b] 
		(5,\x+\tick) node [a] {} -- (5,\x+1-\tick) node [a] {};
	}

\end{tikzpicture}
\endpgfgraphicnamed
\quad \quad 
b) \beginpgfgraphicnamed{comb}
\begin{tikzpicture}
	[b/.style={fill=blue,fill opacity=.3},
	c/.style={fill=black,fill opacity = 1},
	scale=.55]

	\def\n{12}

	\draw (0,0) grid (\n,\n);
	\foreach \x in {0,1,...,\n} \foreach \y in {0,1,...,\n}{
		\filldraw[c, shift={(\x,\y)}] (0,0) circle (1.8mm);
	}

        \fill[b] (-0.5,-0.5) rectangle(\n+.5,0.5);
        \foreach \x in {0,3,...,\n} {
          \fill[b] (\x-0.5,.5) rectangle (\x+0.5,\n+.5);
        }	
\end{tikzpicture}
\endpgfgraphicnamed
\caption{a)~A quantum state of $n$ qudits on a two dimensional grid. The construction is inspired by projected entangled pair states which are important in condensed-matter physics~\cite{Verstraete2008}. 
b)~A ``comb'' of vertices. We consider sets $S$ which contain all of the blue points and any subset of the remaining points. This is chosen so that the resulting set of vertices is always connected. If the grid has $n$ vertices total, then the number of choices for $S$ is $2^{2 n/3-O(n^{1/2})}$.}\label{fig}
\end{figure}

Consider any two-dimensional grid graph with $n$ vertices such as that depicted in \Fref{fig}a, where each vertex (large circle) denotes a qudit with $d = 2^\nu$, where $\nu$ is the degree of that vertex. We encode $\nu$ qubits into each qudit in the obvious way, by just identifying $\C^d \simeq (\C^2)^{\otimes \nu}$. Each edge $e$ corresponds to a two-qubit state $\rho_e$ which is far from separable in the trace distance, namely
\begin{align}
	\inf_{\sigma \in \Sep(e)} \|\rho_e - \sigma\|_\tr \ge \delta \,.
\end{align}
Here $\delta>0$ is a constant and the separability is with respect to the (unique) bipartition of the two-qubit embedded space only, not the larger qudit space. 

The various subsets $S$ of the vertices in the lattice are either connected or disconnected. The disconnected subsets always contain a bipartition with separable states, so for these we have $\Delta(\rho_S) = 0$. For the connected subsets $S$, any bipartition of the nodes contains an edge from $A$ to $A^c$ for all possible choices of $A$. We now show how this implies a constant lower bound on the trace distance to the set $\Sep(A)$. 

The trace distance is contractive for any completely positive trace-preserving (CPTP) linear map, meaning that $\|\cE(M)\|_\tr \le \|M\|_\tr$ for any CPTP $\cE$ and any Hermitian operator $M$\cite{Perez-Garcia2006}. In particular, the partial trace is a CPTP map, so for any bipartition we can choose an edge $e$ between $A$ and $A^c$ and trace over the complement of that edge. We then have
\begin{align}
	\inf_{\sigma \in \Sep(A)} \|\rho_S - \sigma\|_\tr \ge \inf_{\sigma \in \Sep(e)} \|\rho_e - \sigma\|_\tr \ge \delta \,,
\end{align}
where the last inequality follows by assumption on $\rho_e$. Since this is true for all bipartitions of $S$ into $A$ and $A^c$, we have $\Delta(\rho_S) \ge \delta$ for all connected $S$ with $|S|\geq 2$.

If follows that for this class of states $K(\rho) \ge N \delta$ where $N$ is the number of connected subgraphs of the grid with $n$ vertices. Using a comb construction (see \Fref{fig}b), we can show that $N\geq 2^{2n/3-O(n^{1/2})}$. This is exponentially large, so even very small values of $\delta$ would also work.

\subsubsection*{Physical implementation?}
Could these states appear in Nature? For this to happen, we merely need to prepare $O(n)$ independent copies of $\rho_e$ and then group them together. This grouping does not have to be a physical operation, although to make it more convincing it may help to have the four qubits being grouped together in physical proximity. To achieve this, we could store the qubits in the polarization states of photons that are prepared in entangled states. If pairs of photons are created in the middle of each edge and fly away from each other, then after a suitable delay, they will each arrive at the vertices around the same time. When this happens, we obtain the state depicted in \Fref{fig}a, even if momentarily.

In principle, the four photons could interact with some particle at each vertex and transfer their entanglement there. Or they could be trapped in cavities or optical fibers at each site, and thereby create longer-lasting entanglement. The goal of this note is to create the simplest possible thought experiment to refute Kalai's Conjecture C, but variants of that conjecture could be countered with variants of this thought experiment.

Indeed, the advantage of our simple construction is that there's no possibility of correlated noise. If we believe that entangled pairs of photons can be created in spacelike-separated experiments and in labs that are initially uncorrelated, then this permits the state in \Fref{fig}a to be created.

We remark that the 2D nature of the graph is essential. It is easy to see that Kalai's conjecture holds for a 1D chain which has only nearest-neighbor entanglement. In this case, only $\poly(n)$ choices of $S$ will be connected, and thus have $\Delta(\rho_S) > 0$. 

Finally, we note that our construction is inspired by the class of states known as projected entangled pair states, or PEPS~\cite{Verstraete2008}. This class of states is notable precisely because they appear to be relevant to the study of actual physical systems. They provide a variational class that seems to capture many features of interest in the ground states of a vast number of 2-dimensional strongly interacting spin systems. This gives us even more confidence that states such as those depicted above are not just possible to prepare in principle, but in fact have already been extensively studied under realistic laboratory conditions.

\subsubsection*{What about qubits?}
Our example requires that each subsystem have at least dimension 8,
but this does not correspond to anything physical; rather, it can be
thought of as forcing the minimum in \Eref{eq:Delta-def} and the sum in
\Eref{eq:K-def} to respect our coarse-graining of the qubits.

Simply taking pairs of entangled qubits without this coarse-graining
would actually yield a polynomial $K(\rho)$, consistent with Kalai's
Conjecture C.  The only sets $S$ with nonzero $\Delta(\rho_S)$ would
be sets consisting of a single entangled pair of qubits. Since there are
only $n/2$ pairs, we would have $K(\rho) = O(n)$.

What about cluster states on a graph? These also fail. If $S$ is any proper subset of the qubits and $S^c$ is its complement, then every qubit within $S$ that is adjacent to a qubit in $S^c$ will be completely dephased when we trace over $S^c$. Taking $A$ to be a single such qubit would result in a separable state. In this case, we would have $K(\rho) = O(1)$.

We do not believe that the dimension of subsystems is a fundamentally
important distinction; in other words, whatever physical principle is
implicated by Conjecture C should not depend on whether it holds for
$d=2$ or $d=8$.  Nevertheless, in the following section, we describe a
physically plausible route to a state of $n$ {\em qubits} with
exponentially large $K(\rho)$.

\section*{The W state}
In this section, we discuss another counterexample, which turns out to be the well-known multipartite W state (see below for definition).  As with the 2D grid, the state definitely has a large value of $K(\rho)$, although whether it qualifies as a counterexample depends on how confidently we can say it has been, or could be, prepared in experiments.  Define the W state on $n$ qubits as
\ba
\ket{W_n} := \frac{1}{\sqrt{n}}\sum_{k=1}^n \ket{0^{k-1},1,0^{n-k}}.
\ea

The loss-tolerance of $\ket{W_n}$ was previously discussed by~\cite{Stockton2003}, although they did not specifically consider Kalai's
measure of entanglement. One appealing feature of the state $\ket{W_n}$ is that it can be prepared in the electronic spin states of some
kinds of atoms~\cite{Stockton2004}. For example, we can prepare it by first polarizing the atoms, putting them in the
$\ket{0^n}$ state, and then sending in a single long-wavelength photon.
If this photon is absorbed, it will cause exactly one excitation, and
if the wavelength of the photon is long enough, this excitation will
be completely delocalized among all the atoms. 
We remark that some trapped ion experiments have already prepared the
state $\ket{W_8}$ with reasonably high fidelity~\cite{Haffner2005}.
Another possibility, suggested to us by Joe Fitzsimons, is to prepare
a single photon in a single mode and then split into a superposition
over many modes using a diffraction grating.

Now let us evaluate $K(\proj{W_n})$.  If we trace out $n-k$ qubits (leaving $k\ge2$ behind), we are left with the state
\ba 
\rho_k := \tr_{n-k}\proj{W_n} = \frac{n-k}{n}\proj{0^k} + \frac{k}{n}
\proj{W_k}.
\ea
To analyze the entanglement across a bipartition with $j$ and $k-j$
qubits, we observe that $\ket{W_k} = \sqrt{j/k}\ket{W_j}\ket{0^{k-j}}
 + \sqrt{1-j/k}\ket{0^j}\ket{W_{k-j}}$.  Thus, the reduced states are
 supported entirely on the subspaces spanned by
 $\{\ket{0^j},\ket{W_j}\}$ and 
$\{\ket{0^{k-j}},\ket{W_{k-j}}\}$.
In this basis, $\rho_k$ looks like
\ban
\rho_k = \bordermatrix{
&\ket{0^j,0^{k-j}}&\ket{0^j,W_{k-j}} &\ket{W_j,0^{k-j}}&\ket{W_j,W_{k-j}}\\
\ket{0^j,0^{k-j}} & 1-\frac{k}{n} &0&0&0&\hspace{1cm}\\
\ket{0^j,W_{k-j}} &0 & \frac{j}{n} & \frac{\sqrt{j(k-j)}}{n} & 0 \\
\ket{W_j,0^{k-j}} & 0 & \frac{\sqrt{j(k-j)}}{n} & \frac{k-j}{n}& 0\\
\ket{W_j,W_{k-j}} & 0 & 0 & 0 &0 
}
\ean
To analyze the distance from separable states, we can use the PPT
(positive partial transpose) test~\cite{PPT1,PPT2}.  Define the
partial transpose $\Gamma$ by $\ketbra{i,j}{k,l}^\Gamma :=
\ketbra{i,l}{k,j}$ (exchanging the second pair of indices.) 
Since $\Gamma$ preserves the set of separable
states, if $\rho^\Gamma$ is not positive semidefinite, then $\rho$
must be entangled.  To make this quantitative, define the negativity~\cite{VW02}
$\cN(\rho)$ to be the sum of the absolute value of the negative
eigenvalues of $\rho^\Gamma$, or equivalently, 
$$\cN(\rho) := \frac{\|\rho^\Gamma\|_1 - 1}{2}.$$

To translate this into a lower bound on the trace distance from
separability, we observe that $\|X^T\|_1 = \|X\|_1$ for any matrix
$X$.  As a result if $X$ is a matrix on $d\times d$ dimensions, we
have $\|X^\Gamma\|_1 \leq d \|X\|_1$, which follows from convexity, the Schmidt decomposition, and Cauchy-Schwarz. 
%
For any separable $\sigma$, 
\ba \|\rho-\sigma\|_1
\geq \frac{1}{d}\|\rho^\Gamma-\sigma^\Gamma\|_1
\geq \frac{\|\rho^\Gamma\|_1 - \|\sigma^\Gamma\|_1}{d}
\geq \frac{\|\rho^\Gamma\|_1 - 1}{d}
 = \frac{2 \cN(\rho)}{d}.\ea

To analyze the spectrum of $\rho_k^{\Gamma}$, we see that
\ban
\rho_k^\Gamma = \bordermatrix{
&\ket{0^j,0^{k-j}}&\ket{0^j,W_{k-j}} &\ket{W_j,0^{k-j}}&\ket{W_j,W_{k-j}}\\
\ket{0^j,0^{k-j}} & 1-\frac{k}{n} &0&0&\frac{\sqrt{j(k-j)}}{n}&\hspace{1cm}\\
\ket{0^j,W_{k-j}} &0 & \frac{j}{n} & 0 & 0 \\
\ket{W_j,0^{k-j}} & 0 & 0 & \frac{k-j}{n}& 0\\
\ket{W_j,W_{k-j}} & \frac{\sqrt{j(k-j)}}{n} & 0 & 0 &0 
}
\ean
There are three blocks: one with eigenvalue $j/n$, one with eigenvalue $(k-j)/n$ and one with eigenvalues $\frac{n-k}{2n}\pm\sqrt{(\frac{n-k}{2n})^2 + \frac{j(k-j)}{n^2}}$. The negative branch of the square root is the only eigenvalue which is negative. From these observations, we can exactly compute the negativity to be
\begin{align}\label{E:negk}
	\cN(\rho_k) = \frac{1}{2} \left(\frac{k}{n} + \sqrt{\frac{4 j (k-j)}{n}+\frac{(n-k)^2}{n^2}}-1\right)\,.
\end{align}
The minimum in \Eref{E:negk} is achieved when $j=1$, and we obtain the following inequalities
\begin{align}
	\cN(\rho_k) \ge \cN(\rho_k)\bigr\vert_{j=1} = \frac{1}{2} \left(\frac{k}{n}+\sqrt{\frac{k^2}{n^2}+\frac{2 k-4}{n}+1}-1\right)\geq \frac{k}{2 n} \,.
\end{align}
The distance from the separable states is then $\Delta(\rho_k) \ge k/8n$. Summing over $k$ in \Eref{eq:K-def} gives
\begin{align}
	K\bigl(\proj{W_n}\bigr) \ge 
	\sum_{k=2}^{n} \binom{n}{k} \frac{k}{8n} = \frac{1}{16}(2^n-2)\,.
\end{align}

\section*{A variant of Conjecture C}
A different version of Conjecture C is in \cite{Kalai2009}, where the
``censorship conjecture'' was formulated in terms of pure states.
There, an alternate entanglement measure, which we call $\tilde
K(\psi)$ (defined below), was conjectured to be upper-bounded by
$\poly(n)$ for all physically realizable pure states $\ket\psi$.  One
difficulty is that all physically realizable states are likely to be
mixed, although one can imagine extending $\tilde K$ to mixed states
using techniques such as the convex roof extension~\cite{roof}.

However, $\tilde{K}(\psi)$ suffers from a more serious flaw, which is
that it equals zero on all but a measure-zero set of pure states.
Thus, the conjecture is certainly true, but in a way that does not have any
implications for the feasibility of quantum computing.

To explain this, we need to define $\tilde{K}(\rho)$ more precisely.
If $\rho$ is a state on $n$ qudits, then Kalai~\cite{Kalai2009}
defines $\tilde{K}(\rho) = \sum_{A\subseteq [n]} \bigl[-S(\rho^{\vphantom{*}}_A) +
\max_{\rho^*} S(\rho^*_A)\bigr]$, where the maximum is taken over all
$\rho^*_A$ that have the same marginals as $\rho_A$ on all proper
subsets of $A$.
However, if $\rho$ is pure, then generically the only state that
agrees with $\rho$ on all marginal density matrices is $\rho$ itself.
Indeed, this is true with probability 1 even if one considers only
marginals of more than $|A|/2$ subsystems~\cite{JL05}.  
(An example of non-generic states for which this doesn't hold are
($\ket{00000} + e^{i \phi} \ket{11111})/\sqrt2$; here, all five qubits are
needed to figure out the value of $\phi$.)
Since $\rho_A^*=\rho^{\vphantom{*}}_A$ for all $A$ whenever $\rho$ is a generic pure
state, it follows that $\tilde{K}(\psi)=0$ almost everywhere.

\section*{Discussion}
Is entanglement fragile? So-called ``cat'' states like the ``$\phi$'' state above are fragile and random states aren't, but there are many things in between.  Quantum error-correcting code states are known to be robust
to error (indeed, they are designed for this), but constructing them
is a significant challenge that skeptics may doubt on intuitive
grounds, even if quantum mechanics doesn't formally prohibit this.  

In this note, we attempted to argue that in fact robust entanglement is
not very hard to produce. One intuitive explanation for this is
related to the fact that random states are entangled with overwhelming
probability. There simply aren't that many product states out there,
and even the volume of separable states is a small fraction of the set
of all possible density matrices.

And what does this have to say about the original motivation for
Kalai's conjectures, namely, skepticism of fault-tolerant quantum
computing? While our examples certainly do not prove anything about
the feasibility of fault-tolerant quantum computing, we do believe
they point to the difficulty of defining simple physically observable
signatures of quantum computing that could be said to be different
from previously observed physics.

Conjecture C, as we see it, is an attempt at a Sure/Shor separator (as
Scott Aaronson puts it~\cite{SureShor})
that would distinguish states we have definitely already seen from the
sort of states we would find in a quantum computer.  It represents an
admirable attempt to formulate quantum computing skepticism in a rigorous and
testable way.

However, we believe that our counterexamples are significant not
especially because they refute Conjecture C, but because they do so
while side-stepping Kalai's main points about quantum error correction
failing.  More generally, it is significant that it is so hard
to come up with a sensible version of Conjecture C.  In our view, this
is because quantum computers harness phenomena, such as entanglement
and interference, that are already ubiquitous, but merely hard to
control.

The situation is somewhat similar to classical 
computing\footnote{Indeed, the strongest argument against almost any
  conjecture that fault-tolerant quantum computing is impossible is that either the conjecture
  would rule out fault-tolerant classical computing, or that it makes physically
  unjustified distinctions between noise in different bases.}.  A modern
data center exists in a state of matter radically unlike anything ever
seen in pre-industrial times.  But when attempting to quantify this with a
crude observable, then it's hard to come up with anything that wasn't
already seen in much simpler technology, like light bulbs, albeit on vastly different scales.  Our note
can be thought of as showing that Conjecture C refers to a correlation
measure that is high not only for full-scale quantum computers, but
even for the quantum equivalent of light bulbs, i.e.\ technology which is
non-trivial, but by no means complex.

\section*{Acknowledgments}
We are grateful to Joe Fitzsimons, Dick Lipton, Ken Regan and especially Gil Kalai for stimulating discussions. STF was funded by the IARPA MQCO program. AWH was funded by NSF grants 0916400, 0829937, 0803478, DARPA QuEST contract FA9550-09-1-0044 and the IARPA QCS contract. 

\bibliographystyle{titles}
\bibliography{KalaiC}

\end{document}